\documentclass[10pt,conference]{IEEEtran}
\IEEEoverridecommandlockouts
\usepackage{ifpdf}

\ifCLASSINFOpdf
\usepackage[pdftex]{graphicx}
\else
\fi
\usepackage{url}

\usepackage{threeparttable}

\usepackage{pgfplots}
\usepackage{tabularx}
\usepackage{makecell}
\usepackage{multirow}
\usepackage{rotating}
\usepackage{xcolor}

\usepackage[font=small,skip=0.5pt]{caption}

\usepackage{listings}
\makeatletter
\def\lst@makecaption{%
  \def\@captype{table}%
  \@makecaption
}
\makeatother

\begin{document}

%
\title{\huge KGSecConfig: A Knowledge Graph Based Approach for Secured Container Orchestrator Configuration}





\author{\IEEEauthorblockN{
Mubin Ul Haque, M. Mehdi Kholoosi, and
M. Ali Babar}
\IEEEauthorblockA{Centre for Research on Engineering Software Technologies (CREST)}
\IEEEauthorblockA{School of Computer Science, and Engineering, The University of Adelaide, Adelaide, Australia}
\IEEEauthorblockA{Cyber Security Cooperative Research Centre}
}



%


\maketitle
\thispagestyle{plain}
\pagestyle{plain}

\begin{abstract}
Container Orchestrator (CO) is a vital technology for managing clusters of containers, which may form a virtualized infrastructure for developing and operating software systems. Like any other software system, securing CO is critical, but can be quite challenging task due to large number of configurable options. Manual configuration is not only knowledge intensive and time consuming, but also is error prone. For automating security configuration of CO, we propose a novel Knowledge Graph based Security Configuration, KGSecConfig, approach. Our solution leverages keyword and learning models to systematically capture, link, and correlate heterogeneous and multi-vendor configuration space in a unified structure for supporting automation of security configuration of CO. We implement KGSecConfig on Kubernetes, Docker, Azure, and VMWare to build secured configuration knowledge graph. Our evaluation results show 0.98 and 0.94 accuracy for keyword and learning-based secured configuration option and concept extraction, respectively.
We also demonstrate the utilization of the knowledge graph for automated misconfiguration mitigation in a Kubernetes cluster. We assert that our knowledge  graph based approach can help in addressing several challenges, e.g., misconfiguration of security, associated with manually configuring the security of CO. 
\end{abstract}

\textbf{Keywords-} Container, Configuration, Security, Knowledge Graph

%
\IEEEpeerreviewmaketitle

\vspace{- 10 pt}
\section{Introduction}
Container Orchestrator (CO), such as Kubernetes \cite{kubeDoc}, Docker-Swarm \cite{swarmDoc}, Nomad \cite{nomadDoc} enable system administrators to perform multiple tasks, e.g., scaling and interconnecting a large number of containers seamlessly \cite{bernstein2014containers}.
Furthermore, organizations gain a significant benefit in deploying containers in clouds across different production environments \cite{burns2019kubernetes} for rapid software delivery using continuous software engineering.

Despite the reported benefits of using CO, security is one of the key concerns while deploying containers in CO \cite{portworx, shamim2020xi}. 
CO is typically used to manage thousands of containers hosting business critical services, whose security can be compromized if CO is not secured. 


That is why CO security misconfigurations have emerged as one of the biggest concerns of system administrator and software professionals \cite{stackrox}. In a recent survey \cite{redhat-state-kubernetes}, CO security misconfiguration has been mentioned as the top concern (59\%). By exploiting a misconfigured identity management roles of a CO in a Capital One system in 2019 \cite{capitalonebreach}, hackers accessed 30 GB of application data containing valuable financial and personal information of 106 million people. 


If the security of a CO is not appropriately configured, it cannot enforce security policies (e.g., restriction of capabilities, accessibility of root files, privileges to run containers) for the underlying clusters on its own as the default options are not always security-focused \cite{datree}. Therefore, system administrators need to manually perform security configuration tasks, such as identifying and updating the default options, implementing configuration arguments and the options involved in the security policy \cite{manCO}.
Moreover, when deploying containers in a cluster, system administrators may need to maintain diverse configuration files for different purposes, such as deployment (e.g., replica, namespaces), managing credentials (e.g., roles and resources), and observing performances (e.g., audit and log). Each of such configuration files consists of several arguments. For example, a deployment configuration file in Kubernetes may consist of 40 configuration arguments \cite{Kubernetes-Deployments} and more than 100 options. Besides, CO has a distributed architecture constituting different components with a vast configuration space, e.g., arguments, options, default values and types. For example, Kubernetes consists of eight different components, more than 1K configuration arguments across all components, which require manual intervention for configuration \cite{manCO}. Thus, the process of manually identifying and configuring a vast configuration space and different configuration files from the security perspective in CO is effort-intensive, time-consuming, and error-prone \cite{datree, manCO}. 

Therefore, automation is highly desirable for securing the configuration of CO to reduce the required effort and minimize the risk of misconfiguration \cite{shamim2020xi, bose}. However, automation support may need to overcome several challenges, such as configuration data \textit{scatteredness}, \textit{dynamicity}, and \textit{overload} for automating the security configuration task of CO. We provide brief explanation of each of these challenges below to help better understand our research's motivators.


Automation of CO security configuration is a knowledge-intensive task; it is important to know `what' options are required for `which' arguments for each component of a CO. This task is not trivial due to a large number of available configuration options and arguments. In addition, CO is typically integrated with multi-vendor software systems \cite{redhat-state-kubernetes}, such as Continuous Integration and Development or Deployment (CI/CD) tools, e.g., Jenkins; Development and Operations (DevOps) platforms, e.g., Docker; and cloud resource providers, e.g., Azure. Therefore, administrators need to investigate each of the \textit{scattered} tools, platforms, or providers' configuration space for identifying and capturing the secured configuration arguments and options. 

CO security configuration options data may \textit{dynamically} change due to new vulnerabilities, malware, security flaws, and patch releases. For example, the configuration option for the argument \textit{`imagePullPolicy'} in Kubernetes was expected to change after Distributed Denial of Service (DDoS) attack in DataDog \cite{datree}. Hence, to mitigate the rapidly changing threat landscape, configuration space needs to be updated continuously, which requires a frequent manual monitor and search of diverse data sources (e.g., HTML, JSON or XML).



Cyber security information \textit{overload} as the number and the sources of threat advisories are increasing. 
Whilst threat advisories contain in-depth information on how attackers target the existing configuration to compromise software systems, it is a cognitively challenging task to separate relevant from the irrelevant information. 
For example, to obtain the secured configuration options for kube api server \cite{kubeapiserver} (one of the eight constituent components of Kubernetes), relevant documents contain 400 sentences on average, whereas the concepts, i.e., relevant sentences, such as reason, implementation details, are 30 sentences on average. 


There is a need for a unified solution for secured CO configuration, constituting configuration space associated with their secured arguments, options, and concepts of heterogeneous, multi-vendor, and diverse tools, platforms, providers for enabling  automation of secured CO configuration.

There are a few studies that have discussed the best security practices \cite{shamim2020xi} and defects \cite{bose} for Kubernetes. Moreover, Kubernetes usage for monitoring resources and performances had also been reported in \cite{monitorsystems, modak2018techniques}.
However, there is a general lack of solutions for automating CO security configuration. 

To address the above identified problem, we propose to build a KG for Secured Configuration (KGSecConfig) that provides a fundamental support for automatic security configuration of CO. 
KG can provide the unification of configuration space associated with their secured arguments, options, and concepts in terms of entities and relations. 
For example, configuration options, arguments, types, default values, code snippets (e.g., Infrastructure-as-Code (IaC)) represent entities. Relations connect entities using phrases that describe relationship among entities, such as \textit{`is'} or \textit{`has'} relationship. 
Besides, we propose a keyword and learning-based model to address the \textit{dynamicity} and \textit{overload} barriers, which can automatically extract secured configuration knowledge in terms of secured arguments, options, and concepts hidden in large documents.
These types of knowledge are populated in the KG for organized structure, so that the encapsulated knowledge can be  utilized for automating the security configuration of CO.
Our KGSecConfig can capture and link the configuration knowledge through real-time monitoring of multiple, diverse, and heterogeneous configuration space from different data sources to reflect the intricate and evolving security threats. Our KGSecConfig can be utilized for various downstream configuration tasks, such as automated misconfiguration mitigation and interpretation of secured configuration options which can increase the understanding of the reason of enabling/disabling particular arguments. Furthermore, our KGSecConfig can provide a way for visualizing configuration arguments where default values are not security-focused and can help administrators to prioritize their efforts in configuration task.    

Our paper makes the following three main contributions:
\begin{itemize}
    \item We are the first, to the best of our knowledge, to propose a keyword and learning-based approach for CO to capture, store and correlate configuration knowledge automatically.
    \item We build a secured configuration knowledge graph using KGSecConfig on Kubernetes \cite{kubeDoc}, Docker \cite{DockerDoc}, Azure \cite{AzureDoc}, and VMWare \cite{VMWareDoc}.
    \item Our evaluation shows high accuracy to build the KGSecConfig. Moreover, we demonstrated the effectiveness of KGSecConfig in automated mitigation of misconfiguration of a Kubernetes cluster. 
\end{itemize}


The remainder of our paper proceeds as follows. Section \ref{section: background} describes the related work. Section \ref{method} and \ref{sec: implementation} discuss the approach and implementation. The process of the evaluation and results are discussed in Section \ref{sec: results}. We report the implications and limitations in Sections \ref{sec: discussion} and \ref{sec: threats}. Finally, our paper concludes in Section \ref{sec: conclussion} with some future directions.

\vspace{- 2 pt}
\section{Related Work}
\vspace{- 2 pt}
\label{section: background}
Our research is related to the prior studies that have investigated the security of CO, the configuration studies in software engineering, and cyber security KG. 
\vspace{- 3 pt}
\subsection{Security in Container Orchestrator}
\vspace{- 3 pt}
Shamim et al. \cite{shamim2020xi} reported a study of grey literature, e.g., blog posts and tutorials to identify security best practices for Kubernetes. They stated secure usage of Kubernetes requires the implementation of security practices applicable to multiple components within Kubernetes, such as containers \cite{containerss} and pods \cite{pods}. They also advocated for a need of a deep understanding of Kubernetes configurations to implement security practices. Bose et al. \cite{bose} conducted a study on open-source software repositories, e.g., GitLab, and identified commits for updating security-related defects in Kubernetes manifests \cite{shamim2020xi}. They applied closed coding \cite{closecode}, a qualitative analysis technique, on the collected commits to determine commits related to a security defect. Moreover, there were studies on the usage of Kubernetes in creating monitoring systems \cite{monitorsystems} and comparison of performance and resource management \cite{modak2018techniques}. However, our research is different to these studies as we aim to develop an approach which can automatically capture, link, and encapsulate the configuration knowledge of diverse configuration space in a unified KG. which can be used for secured configuration of CO.  
\vspace{- 5 pt}
\subsection{Configuration in Software Engineering}
\vspace{- 3 pt}
Sayagh and Hassan \cite{sayagh} identified the appropriate options to configuration related user questions by mining already answered configuration  questions on online forums. Jin et al. \cite{jin} and Wen et al. \cite{wen} identified the appropriate options to a configuration-related question by extracting options whose option names are textually similar to the new user question. Xia et al. \cite{xinxia} predicted if a bug report is related to configuration or not. However, the study \cite{sayagh} focused on the already answered question; thus, there would be a long trail of less discussed configuration options. Moreover, all the configuration options might not be discussed on online forums, and the answered options might not be reliable, as shown by prior studies \cite{sworna, verdi}. Besides, given a bug report, the study \cite{xinxia} determined whether the bug report was a configuration or not, and the studies \cite{jin, wen} extracted configuration options from the bug reports. However, our research goal is different as our motivation is not only to extract configuration options but also realize and extract configuration concepts (e.g., reason, implementation details). We specifically intend to encapsulate such extracted configuration knowledge in a structured way in the unified knowledge graph. This unified knowledge graph allows us to correlate the vast configuration space of diverse CO software systems and, afterward, enables us to automate the secured configuration and compliance.
\vspace{- 1 pt}
\subsection{Cyber Security Knowledge Graph}
\vspace{- 2 pt}
Rastogi et al. \cite{malkg} proposed a KG for predicting missing information for malware. Pingle et al. \cite{pingle} proposed a system, RelExt, that would detect relationships and create semantic triples over cyber security text using a deep-learning approach. Piplai et al. \cite{piplai} extracted the information from malware After Action Reports (AAR) that can be merged to create a KG using RelExt. Mendsaikhan et al. \cite{mendsaikhan} identified the significance of the cyber security text
using the deep-learning approach. Whilst the prior studies focused on detecting cyber security text, e.g., attack patterns, vulnerabilities, malware, our study, in contrast, aims to build a KG automatically for secured configuration knowledge for CO.


In summary, our research, to the best of our knowledge, has provided a unique solution, KGSecConfig, for building a secured configuration knowledge graph automatically for diverse software systems in CO. Our KGSecConfig is expected to minimize the manual efforts required to extract and utilize configuration knowledge for securing CO configuration. 



\vspace{- 1 pt}
\section{Approach}
\label{method}
\begin{figure}[]
    \centering
    \includegraphics[scale=0.4]{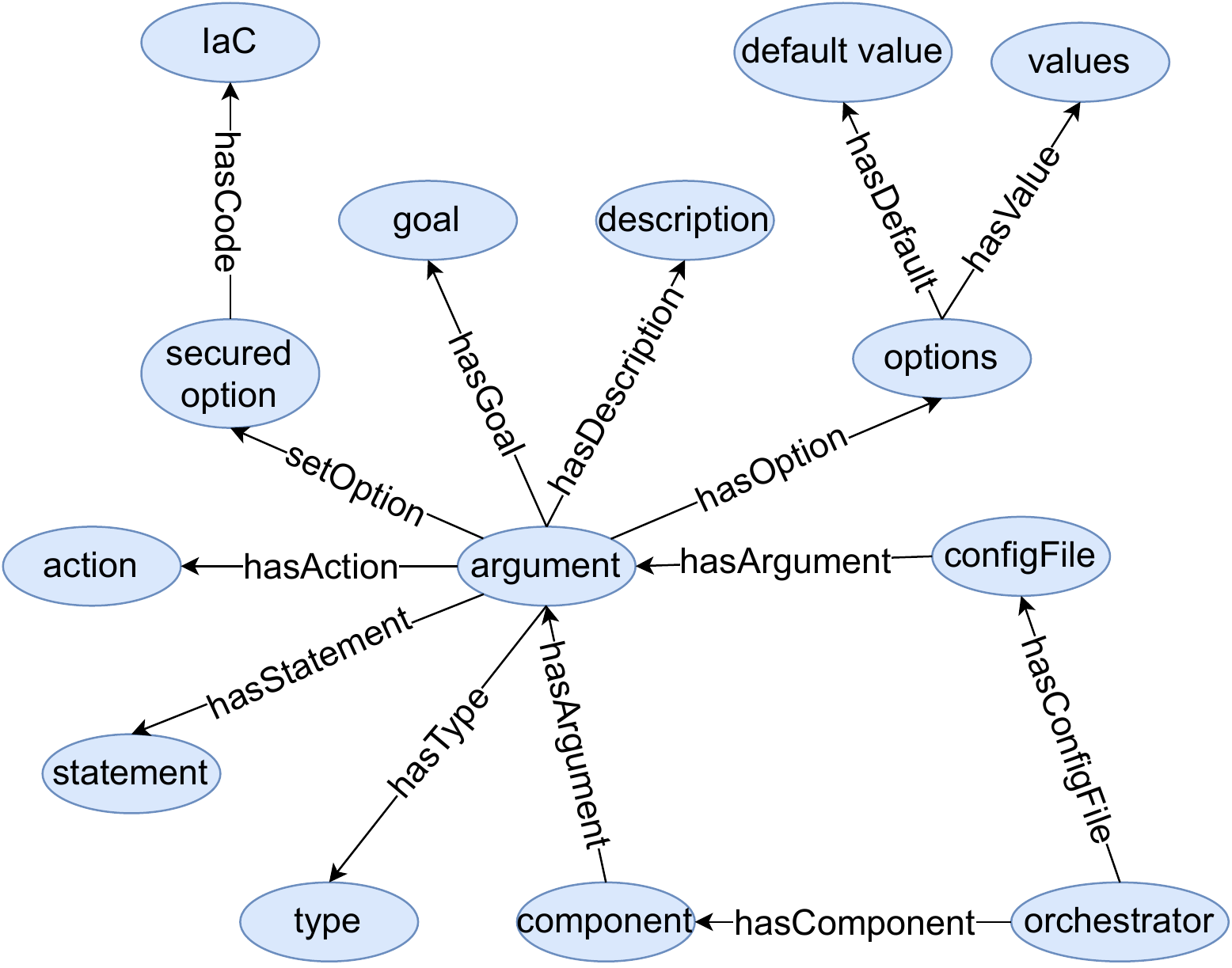}
    \caption{Conceptual schema for configuration knowledge}
    \label{fig:relations}
\end{figure}
\begin{figure}[]
  \centering
  \includegraphics[scale=0.5]{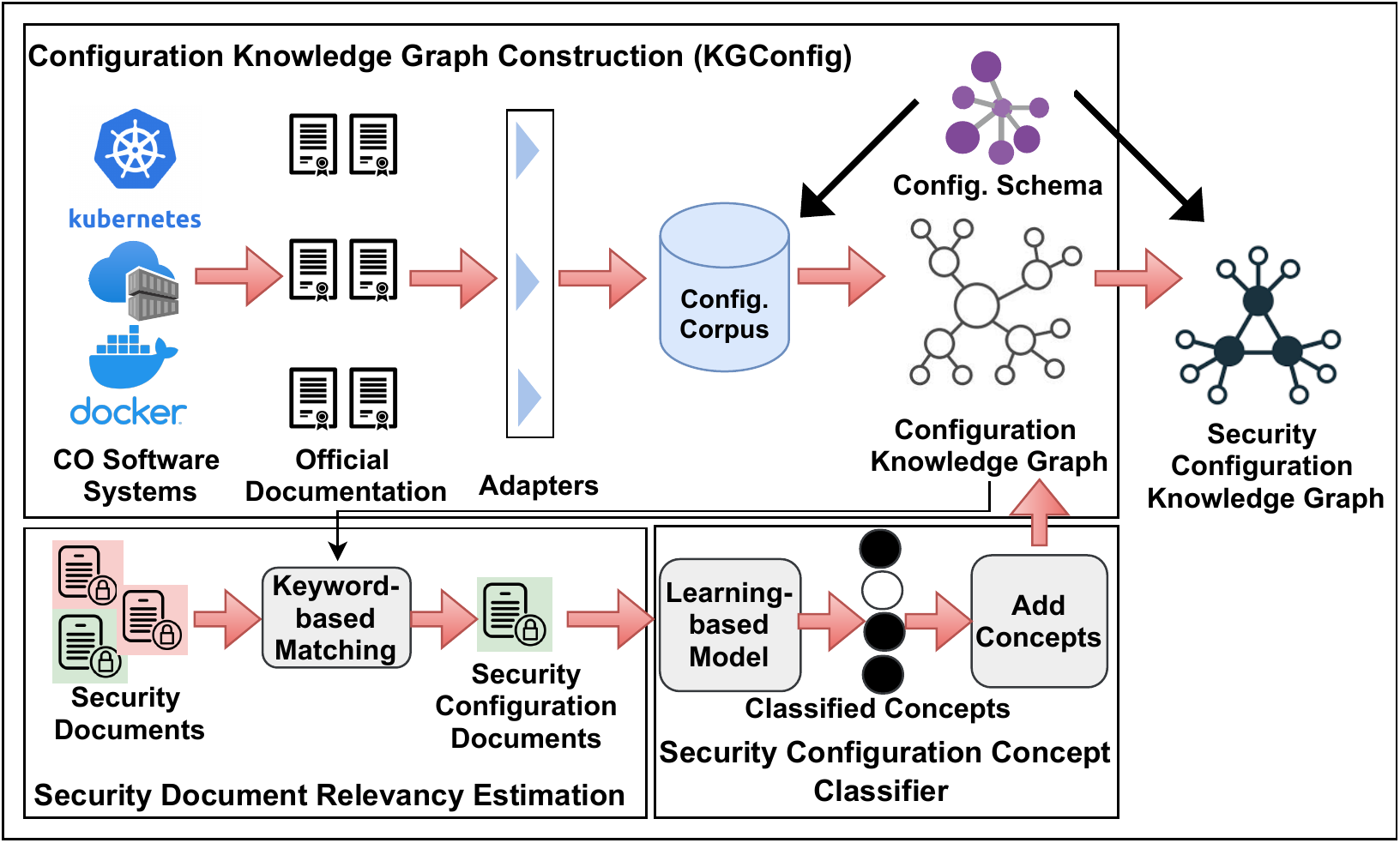}
  \caption{ Approach for constructing KGSecConfig}
  \label{fig: pipeline}
\end{figure}
We propose a Knowledge Graph (KG) based approach to overcome the above-mentioned barrier, i.e., data \textit{scatteredness}, \textit{dynamicity} and \textit{overload}, to automating the configuration of CO. The conceptual schema of our KGSecConfig is shown in Fig. \ref{fig:relations}; and Fig. \ref{fig: pipeline} shows the different stages of envisioned process for automatically constructing KGSecConfig. In particular, we present how KGSecConfig can be constructed from the natural language text sources by retrieving the entities and relations to form the triples \cite{pingle, piplai} of KGSecConfig. Our approach contains three main modules named \textit{Configuration Knowledge Graph (KGConfig) Construction}, \textit{Security Document Relevancy Estimation},  and \textit{Security Configuration Concept Classifier}. The details of each module are described as follows.
\subsection{Configuration Knowledge Graph (KGConfig) Construction}\label{sec: CKG}
\begin{figure}[]
  \centering
  \includegraphics[scale = 0.35]{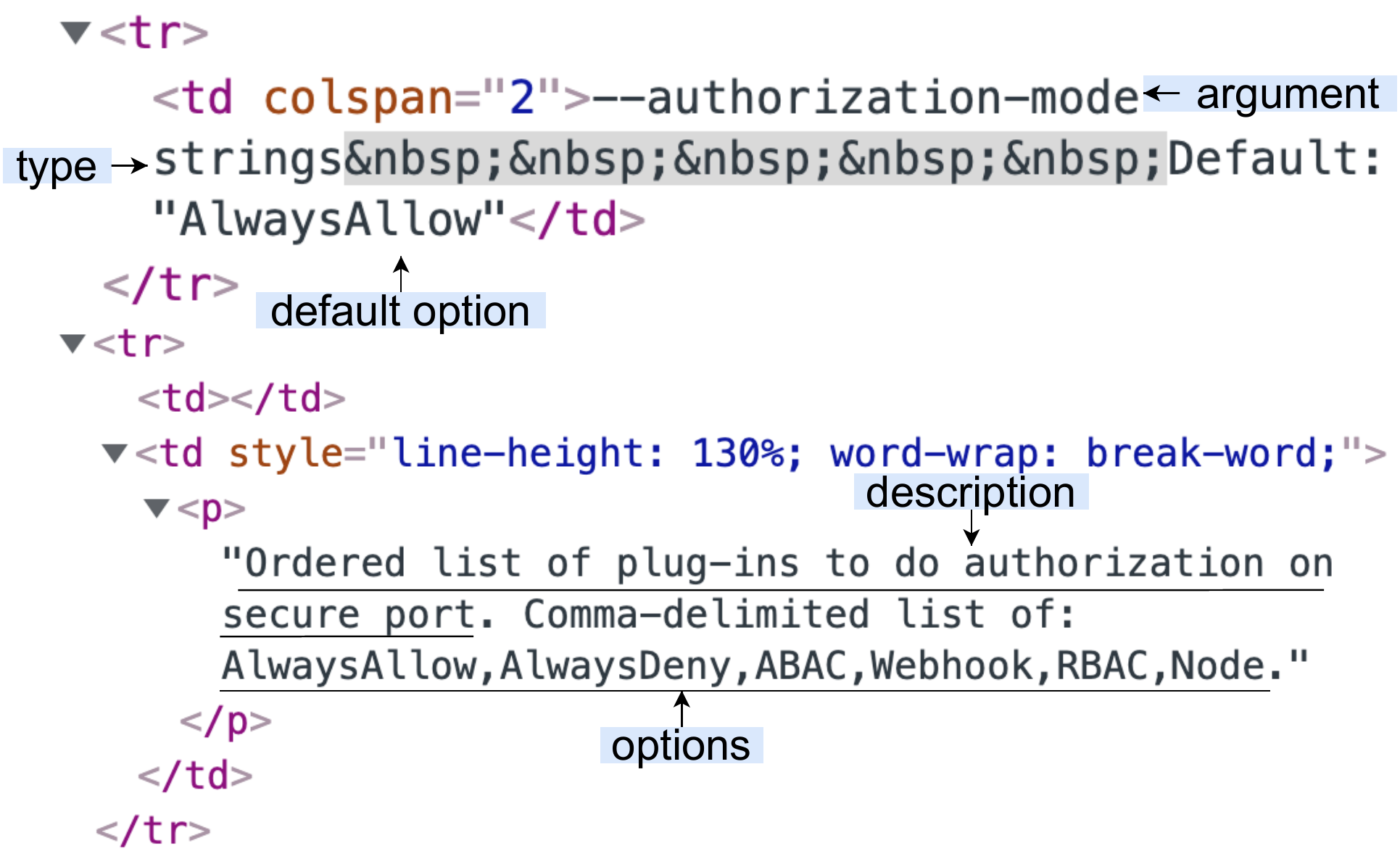}
  \caption{Example of configuration space extraction}
  \label{fig: example1}
\end{figure}

\begin{figure}[]
    \centering
    \includegraphics[scale=0.40]{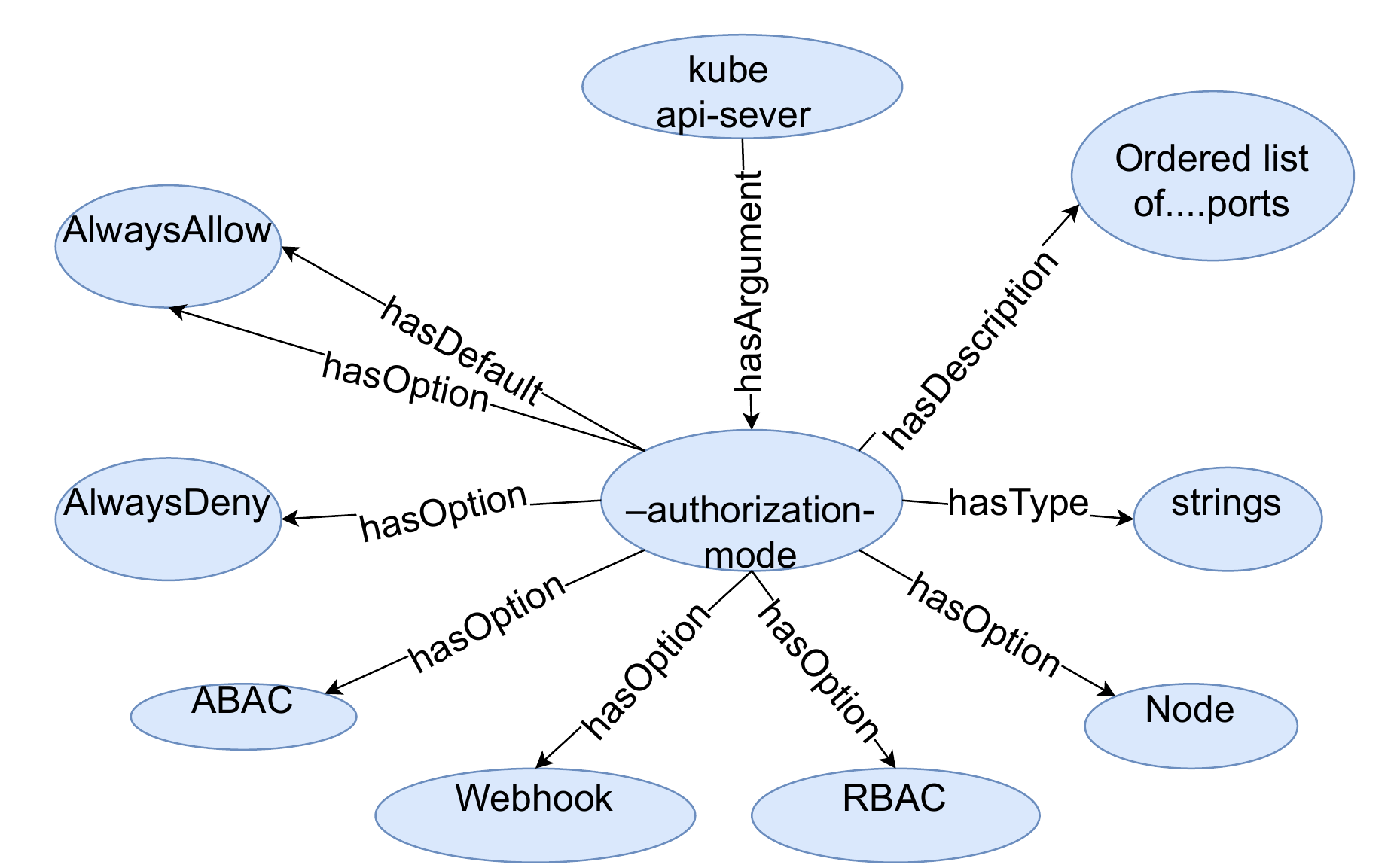}
    \caption{Triples corresponding to Fig. \ref{fig: example1}}
    \label{fig:table1}
\end{figure}

\vspace{1 pt}
The purpose of this module is to create a Knowledge Graph for Configuration (KGConfig) of heterogeneous platforms, cloud resource providers, and tools used in CO to mitigate data \textit{scatteredness}. We encapsulated configuration arguments, options, types, default values, and descriptions. Configuration arguments and options are indispensable knowledge for performing the configuration task (e.g., \textit{`RBAC'} is an option in \textit{`--authorization-mode'} argument for configuring kube api-server). Types and default values are necessary knowledge to verify whether the used types (e.g., strings for \textit{`--audit-policy-file'} argument, or integer for \textit{`MinReadySeconds'} argument) and values in the configuration file (e.g., Kubernetes manifest file) are aligned with the required setting. Descriptions are essential knowledge for discovering and understanding the functionality of the configuration argument. In this module, we used different adapters to extract the configuration information, e.g., arguments, options, descriptions, and default options, from official documentation of diverse orchestrators, tools, platforms and built a configuration corpus with the extracted information. Then we applied our schema as shown in Fig. \ref{fig:relations} to the configuration corpus to develop KGConfig. We need different adapters (e.g., data scrapers) since the documentation content is distributed in diverse formats such HTML, XML, or JSON. One example of extracting configuration information from documentation and constructing triples to populate KGConfig is illustrated in Fig. \ref{fig: example1} and Fig. \ref{fig:table1}. The extracted arguments, options, types, and default options are considered as the entities and we defined \textit{`has'} relationship, e.g., \textit{`hasArgument'}, \textit{`hasOption'}, \textit{`hasType'}, \textit{`hasDefault'} and \textit{`hasDescription'} among the entities. These entities and relations are important for identifying the configuration syntax and used to formulate keyword-based rules for estimating the relevancy of security documents with configuration (Section \ref{sec: SDRE}). In summary, the input of this module is a set of official documentations, and the output is a KGConfig consisting of configuration entities and corresponding relations.  

\subsection{Security Document Relevancy Estimation (SDRE)} \label{sec: SDRE} 
\begin{figure}[]
  \centering
  
  \includegraphics[width=\linewidth]{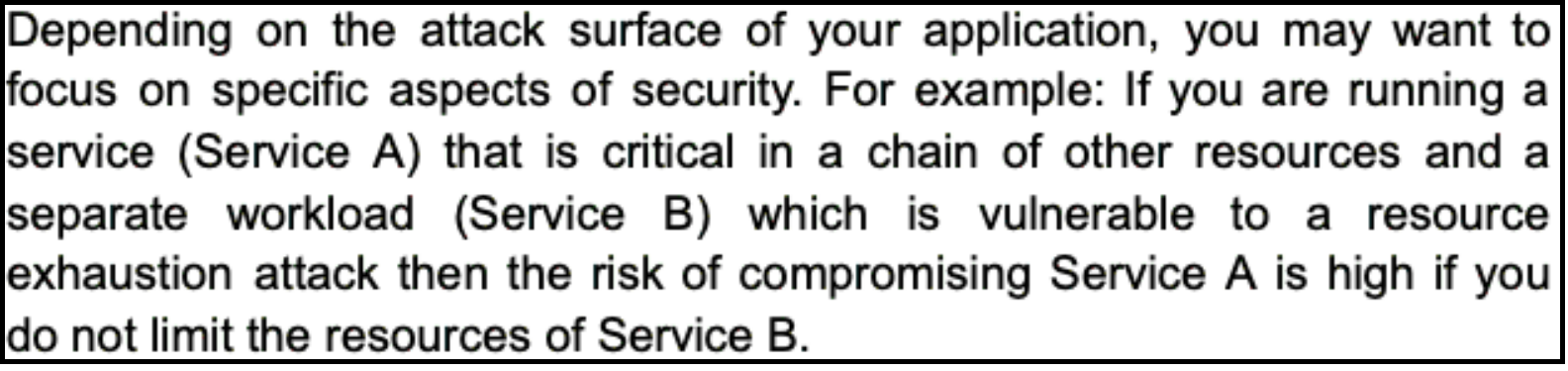}
  \caption{Example of a non-configuration document}
  \label{fig: nonConfig}
\end{figure}

\begin{figure}[]
  \centering
  
  \includegraphics[width=\linewidth]{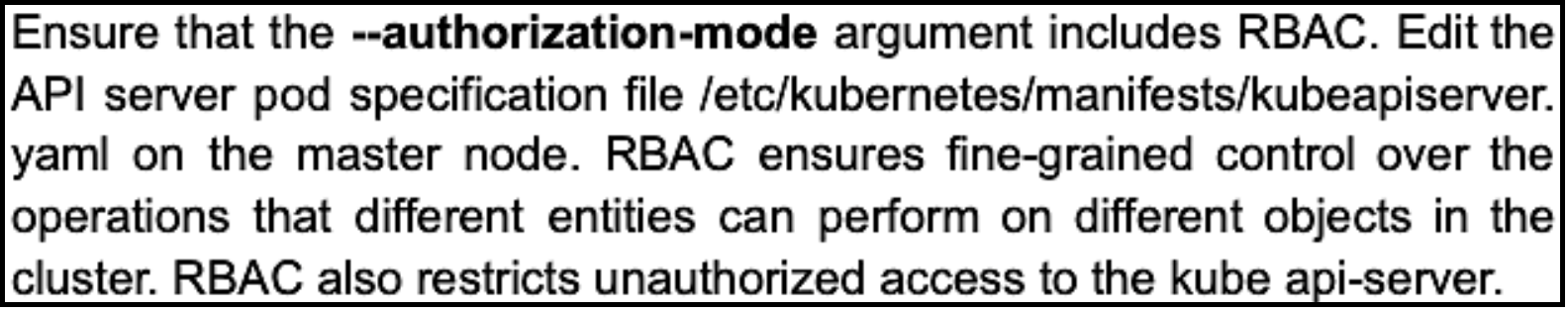}
  \caption{Example of a configuration document}
  \label{fig: ConfigE}
\end{figure}

The purpose of this module is to filter out security documents from configuration point of view to mitigate data \textit{dynamicity} barrier. Our KGSecConfig extracts security configuration information from publicly available security information-sharing sources. While security information-sharing sources (e.g., National Vulnerability Database (NVD), Security Tool White Paper, Security Bulletin) have grown in popularity \cite{mendsaikhan}, the amount of shared security information has also increased tremendously. This security information is shared in text documents and all the documents may not be relevant to configuration. Fig. \ref{fig: nonConfig} and Fig. \ref{fig: ConfigE} show the examples of two excerpts from Kubernetes Blog \cite{kubeblog}, where a document is relevant to configuration and other is not relevant to configuration. In Fig. \ref{fig: ConfigE}, \textbf{bold} text (e.g., configuration argument) represents the reason of a sentence being relevant to configuration. In particular, we labeled a sentence that is relevant with configuration if any word in the sentence is matched with the extracted configuration argument, which is obtained from the previous step (Section \ref{sec: CKG}). In the same way, we considered a security document is relevant to configuration if there is any sentence being labelled as configuration. We used configuration argument and its associated entities as the keyword for estimating the relevancy of security documents to configuration. In summary, the input of this module is a set of security documents and the output is a set of security configuration documents.


\vspace{- 5 pt}
\subsection{Security Configuration Concept Classifier (SCCC)} \label{sec: SCCC} The purpose of this module is to realize and extract configuration concepts from security configuration documents to mitigate data \textit{overload} barrier.
We proposed four main classes for security configuration concepts as (i) \textit{`statement'} concept describes to perform a security configuration task (e.g., Ensure that the \textit{`--authorization-mode'} argument includes \textit{`RBAC'}); (ii) \textit{`goal'} concept describes the reason behind the security configuration task (e.g., \textit{`RBAC'} ensures fine-grained control over the operations...cluster, \textit{`RBAC'} restricts unauthorized access...server); (iii) \textit{`action'} concept describes the actionable steps to implement the task (e.g., Edit the API server pod specification file...node); and (iv) \textit{`other'} concept describes the rest of sentences in the document (e.g., In this article, we will take a deep dive into key Kubernetes security configurations and the recommended best practices). 

A secured configuration \textit{`statement'} concept is essential to correlate statements with actionable steps in the knowledge graph. For example, Security Announcement Kubernetes \cite{kubernetes-security-announce} did not provide the actionable steps to mitigate the security issues due to CVE-2020-8557, however, it mentioned the statement (e.g., force containers to drop CAP-DAC-OVERRIDE capabilities in PodSecurityPolicy). In this regard, storing a statement in the knowledge graph can help find actionable steps for implementing secured configuration tasks. Moreover, \textit{`goal'} concept is essential to demonstrate `why' a particular option should be used or not for security purposes. Dietrich et al. described system administrators suffer from a lack of knowledge in configuration, which is one of the primary reasons for misconfiguration \cite{dietrich2018investigating}. Therefore, the \textit{`goal'} concept provides a way to interpret the cause of a particular configuration, which can be influential in broadening the configuration knowledge. The \textit{`action'} concept provides the required low-level implementation details, which are necessary for the machine execution of configuration tasks. Besides, \textit{`other'} concept is necessary to filter out irrelevant sentences from the security configuration document.


  

We leveraged learning-based techniques to classify the sentences in security configuration documents automatically. Learning-based techniques are used since they are able to independently adapt new data for their learning ability from the previous computation, e.g., historical data for producing repeatable and data-driven decisions \cite{puminer, zhao2017hdskg, vulners2}. We designed the problem as a multi-class supervised text classification problem from a learning perspective. 
The components of building a learning model include- text \textbf{pre-processing}, \textbf{model selection}, \textbf{model building}, and \textbf{prediction}. 
\textbf{Pre-processing.} security configuration document may contain noise (e.g., punctuation, stop-words), which can make the learning model overfit \cite{puminer}, \cite{kao2007natural}. Therefore, we used the state-of-the-art approaches \cite{vijayarani2015preprocessing} for pre-processing the sentences, e.g., removal of noises, stop-words, lower casing, and lemmatization. Lemmatization is used to minimize the inflectional and derivational forms of a word to a standard base form \cite{zhao2017hdskg}. We left the configuration argument and options intact, e.g., \textit{`--anonymous-auth'}, to preserve the configuration syntax.  \textbf{Model Selection.} we used the pre-processed text to perform stratified k-fold cross-validation. Stratification ensures the ratio of each input source is kept throughout the cross-validation \cite{puminer}, avoiding different data distribution of the folds. Our model selection component has two steps as (i) feature engineering, (ii) model training and validation.  Feature engineering is the process where textual data is transformed into features to improve the performance of the learning models \cite{chadni}. In the model training and evaluation steps, (k-1) folds are used for feature engineering and training a model, while the remaining one is used for validation. The validation performance of a model is the average of k runs. The model configurations with the highest performance metric would be selected as the optimal classifier for the following model building process.  \textbf{Model Building.} the model building process uses the pre-processed data to generate a feature model based on the identified feature configuration. The feature model has been saved to transform the data for future prediction. \textbf{Prediction.} the prediction process is used for both testing the trained model and classifying the new sentences. In this process, the sentences are first pre-processed and then transformed to a feature set using the saved feature model. Finally, the feature set is used by the saved trained model to determine the class of the sentences.

Our SCCC module will reduce the manual analysis required for realizing concepts from security configuration documents. In summary, the input of this module is a security configuration document and the output is the sentences classified according to the four configuration concepts. 

\begin{figure}[]
    \centering
    \includegraphics[scale=0.38]{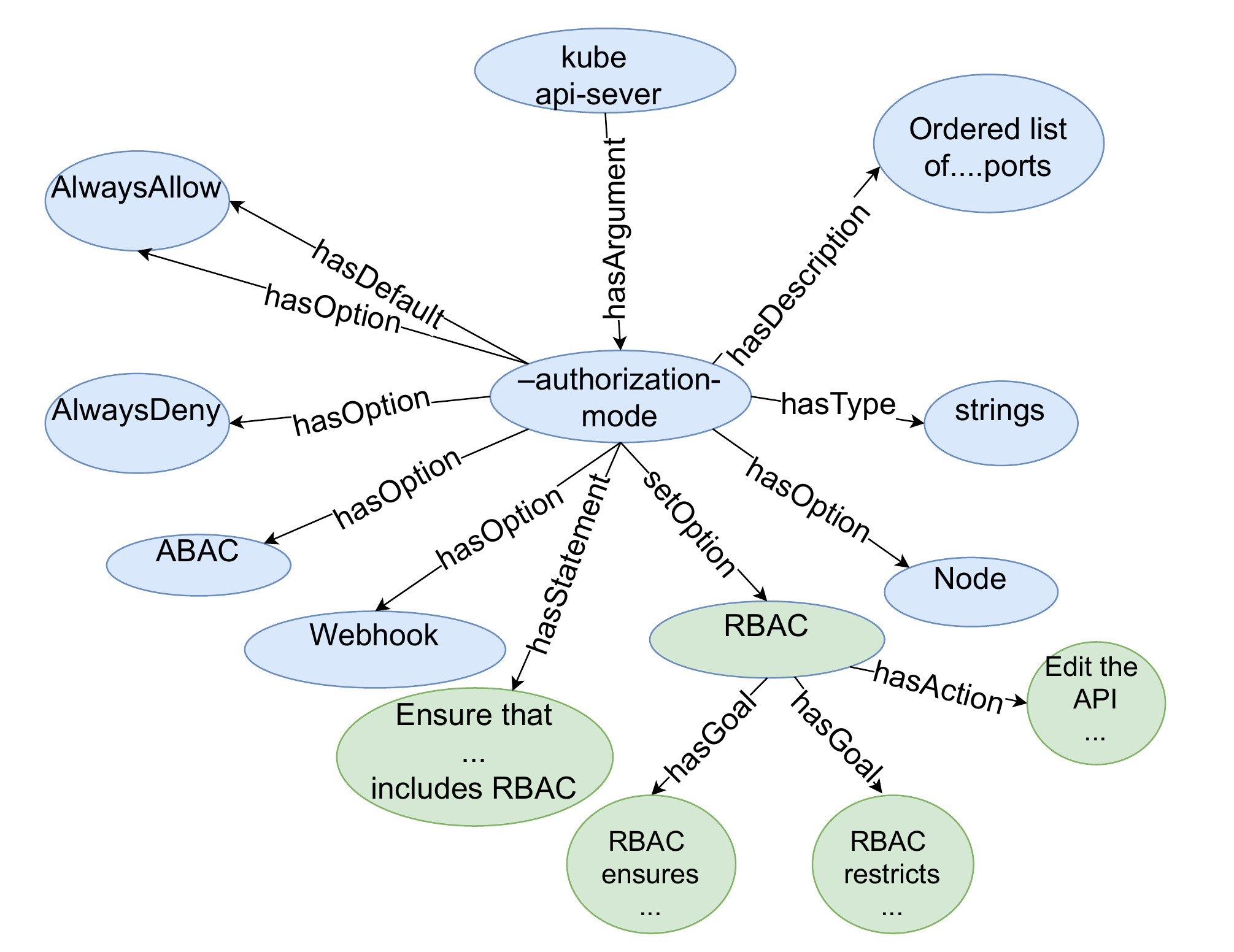}
    \caption{Configuration concepts added to KGConfig (Fig. \ref{fig:table1}) from the text of Fig. \ref{fig: ConfigE}}
    \label{fig:concepts}
\end{figure}

Once we identified the classified sentences, we considered all the concepts except \textit{`other'} concept for updating the initial KGConfig. We used graph-traversal algorithm \cite{bfs} to locate the configuration argument and then added concepts as entities with \textit{`has'} relationship as described in Fig. \ref{fig: pipeline} to construct a secure configuration knowledge graph. Moreover, we identified the required options for security configuration mentioned in the concepts by a rule-based approach. The rules are formulated from the available options associated with the particular arguments that had been initially encapsulated in the KGConfig. Fig. \ref{fig:concepts} shows an example to construct a secure configuration knowledge graph from the identified concepts. 

To mitigate data \textit{scatteredness} barrier, we proposed a configuration schema which is used to capture the configuration arguments, descriptions, default, and available options from diverse, heterogeneous, and multi-vendor tools documentation using entities and relations. While the schema can be used for instantiating KG, it can also map data of different formats (e.g., structured, unstructured, or semi-structured) from multiple sources into a common structure (Section \ref{sec: CKG}). Secondly, we proposed a keyword-based configuration knowledge localization method (Section \ref{sec: SDRE}) to overcome data \textit{dynamicity} barrier. Thirdly, to mitigate the data \textit{overload} barrier, we proposed a learning-based model (Section \ref{sec: SCCC}). The proposed learning-based model can automatically adopt new data from diverse sources to keep the configuration updated for mitigating security incidents.

\vspace{- 7 pt}
\section{Implementation}
\vspace{- 4 pt}
\label{sec: implementation}
We applied our knowledge graph construction approach to Kubernetes \cite{kubeDoc} (container orchestrator), Docker \cite{DockerDoc} (DevOps platform), Azure \cite{AzureDoc} (cloud resource provider), and VMware \cite{VMWareDoc} (infrastructure manager) which can be integrated with Kubernetes \cite{kubecase}. We selected the above four software systems due to their large-scale adoption for integrated orchestration service in a software-defined network \cite{kubecase}. 

We used Beautiful Soup \cite{BeautifulSoup}, a Python library for parsing HTML and XML documents to crawl the web pages of official documentation for knowledge extraction. In addition, NLTK \cite{NLTK}, a Python library for text processing, was used to implement the text pre-processing in descriptive knowledge extraction. Beautiful Soup and NLTK were widely used in software engineering studies involving information retrieval process \cite{liu2019generating, hu2018summarizing}.

We leveraged following security information-sharing sources for secured configuration knowledge extraction.  
\begin{itemize}

    \item \textbf{Security Announcement Kubernetes:} Official site \cite{kubernetes-security-announce} for announcing security adversaries discovered or reported for Kubernetes and its integrated software systems, e.g., Docker or Azure. This site is maintained by Kubernetes security experts and recommended for the Kubernetes users for security task information \cite{kubeDoc}.   
    \item \textbf{Internet Artifacts:} Shamim et al. proposed a taxonomy for secured installation of Kubernetes clusters and shared a dataset of 101 blog posts \cite{shamim2020xi} from internet. Researchers have acknowledged the value of internet artifacts in deriving security tasks in various domains, such as DevOps \cite{rahman2016softwar}, CI/CD \cite{rahman2015synthesizing}, and testing \cite{garousi2018smells}.  
    \item \textbf{Security Tool White Paper:} Security tools, such as SYSDIG \cite{sysdig} release white papers to secure deployment of container workloads and contain valuable information. The previous researches showed the importance of analyzing white papers for realizing security tasks \cite{islam2019multi}. We used Vulners database \cite{vulners}, a rich source of security task information for diverse and heterogeneous software systems from more than 50 security tools. Vulners database was also used by other researches \cite{vulners2, vulners1, vulners3}. 
\end{itemize}
 
\begin{table}[]
\centering

\caption{Description of data set}
\label{tab: datasets}
\scalebox{0.8}{
\begin{tabular}{|c|c|c|c|}
\hline
Source         & Official & Internet  & Vulners \\ \hline
Security-related Document          & 42  & 101 & 2187 \\ \hline
Security Configuration Document  & 39  & 91  & 205  \\ \hline
\end{tabular}
}
\end{table}
Table \ref{tab: datasets} shows the distribution of the documents related to security and configuration. We used our 5,172 arguments as a keyword to search and identify security configuration-related documents. The search was performed in August 2021. 

We required a labelled dataset for the supervised learning model to extract configuration concepts from the security configuration documents. To the best of our knowledge, there is no labelled dataset for security configuration concept classification. Manual labelling of data is time-consuming and labour-intensive \cite{zhao2017hdskg}. Therefore, we randomly selected 3,300 sentences from the security configuration documents for manual labelling. Two authors manually and independently label the sentences based on our four configuration concepts (Section \ref{sec: SCCC}). We used Cohen's Kappa \cite{cohenkappa} to measure the agreement between two labelers. Cohen's Kappa is used since the same two labelers had rated the set of selected sentences, and also used in software engineering studies to report the agreement \cite{332818,knowhow}. We obtained the Kappa value of 0.7, indicating substantial agreement \cite{cohenkappa}. We used 3,032 sentences which were agreed upon by both labelers for training the  model to reduce the labeling bias.       

Five traditional machine learning classifiers, Logistic Regression (LR), Naive Bayesian (NB), Support Vector Machines (SVM), Random Forest (RF), and Extreme Gradient Boosting (XGB), were selected for learning-based models. Those classifiers were chosen due to the common practice in the literature \cite{puminer-15,puminer-47,puminer-48} and real data science competition (e.g., Kaggle \cite{kaggle}). The first three (e.g., LR, NB, SVM) are single models, whereas the rest two (RF, XGB) are ensemble models \cite{puminer}. In addition, we considered Term Frequency-Inverse Document Frequency (TF-IDF)-based word level, character level, and combination of word and character level to obtain relevant features. We also considered NLP features, such as word count, character, noun, verb, adjective, adverb, pronoun, and word density. These features are common and also used in the previous studies of software engineering studies that involve classification process \cite{xinxia, puminer, puminer-15}. To select optimal traditional ML models, we applied Bayesian optimization \cite{bush1} using hyperopt library \cite{bush2}. We chose bayesian optimisation due to its robustness against noisy objective function evaluations \cite{bush3}. We utilised the average Matthews Correction Coefficient (MCC) of 10-fold cross-validation with stratified sampling \cite{puminer} and early stopping criteria to select the optimal hyper parameters. MCC was used to select the optimal model since MCC explicitly considers all classes and is proclaimed as the best metric for error consideration by the prior study \cite{bush4}. Besides, we implemented our models using scikit-learn \cite{scikit-learn} and gensim \cite{gensim}. 

After concept extraction from the corpus, we used the Breadth-First-Search (BFS) \cite{bfs} algorithm to identify the configuration argument and then update the KGConfig. BFS is a traditional graph traversal algorithm and is also commonly used in prior knowledge graph research \cite{bfskg}.

\section{Results and Evaluation}
Our KGSecConfig built upon Kubernetes, Docker, Azure, and VMWare consists of 5,172 arguments, 2,774 options, and 23,177 descriptions. We obtained 1,463 argument, 1,793 options, where 984 arguments are not secured by default, and 479 arguments are secured by default.

We conducted a series of experiments to evaluate the effectiveness and feasibility of our KGSecConfig by answering the following Research Questions (RQs).
\begin{itemize}
    \item \textbf{RQ1}: \textit{How effective is a keyword-based method for estimating relevancy of security documents to configuration?} Our keyword-based method considers configuration arguments with the exact syntax as mentioned in the software systems. The answer to RQ1 will inform whether the method returns the accurate configuration documents. Accurate configuration documents are important since they are the primary source of secure options and their implementation details. 
    \item \textbf{RQ2:} \textit{How effective is our concept classifier module based on machine-learning?} Our classifier module aims to classify sentences of the security configuration documents to identify concepts. The answer to RQ2 will reveal how accurately the module predicts in terms of classification. An accurate model is important since the predicted concepts will be further used to build KGSecConfig on top of KGConfig by adding concepts. Besides, it will help organizations to select suitable models in their orchestration service in a software-defined network.
    \item \textbf{RQ3:} \textit{What is the intrinsic quality of the knowledge captured in KGSecConfig?} Our KGSecConfig aims to capture, link, and correlate the configuration knowledge of multi-vendor software systems from diverse data sources, which can be further utilized for securing the CO configuration automatically. The answer to RQ3 will reveal the quality, i.e., whether the captured knowledge precisely convey the completeness and correctness of the configuration knowledge. Ascertaining the completeness and correctness is crucial for determine KGSecConfig's purpose for various downstream utilization.
    
    
\end{itemize}
\label{sec: results}
\vspace{- 5 pt}
\subsection{Protocol for answering RQs}

\subsubsection{RQ1} We adopted a sampling method \cite{332818} similar to the prior studies \cite{332818_1,332818_2} to ensure that the ratios observed in the sample are generalized to the population within a specific confidence interval at a certain confidence level. The required sample size is 385 for a confidence interval of 5 at 95\% confidence level. Therefore we randomly selected 385 security documents and identified how accurately the keyword-based method could select configuration-related documents. 
\begin{table}[]
\centering
\caption{Evaluation metrics}
\label{tab: evaluationMetrcis}
\scalebox{0.8}{
\begin{tabular}{|c|c|}
\hline
\multicolumn{1}{|c|}{Metric}                                                     & \multicolumn{1}{c|}{Description}                                                                                                                                                              \\ \hline
Accuracy                                                                   & Percentage of correctly classified instances.                                                                                                                                                 \\ \hline
Recall                                                                    & The correctly identified proportion of positive instances.                                                                                                                                    \\ \hline
Precision                                                                        & \begin{tabular}[c]{@{}l@{}}The percentage of the detected positive instances that \\ were correct.\end{tabular}                                                                               \\ \hline
f1-score                                                                         & The harmonic mean of recall and precision.                                                                                                                                                     \\ \hline
\begin{tabular}[c]{@{}l@{}}Matthews correlation \\ coefficient (MCC)\end{tabular} & \begin{tabular}[c]{@{}l@{}}It is correlation coefficient that depicts the performance \\ of the classifier by considering all four dimensions in \\ the confusion metric.\end{tabular} \\ \hline
\end{tabular}
}

\end{table}

\subsubsection{RQ2} We used our manually curated dataset of 3,032 sentences (\textit{`statement'} concept 790 sentences, \textit{`goal'} concept 860 sentences, \textit{`action'} concept 751 sentences, and \textit{`other'} concept 631 sentences) to evaluate the ML-models. Besides, we used the evaluation metrics defined in Table \ref{tab: evaluationMetrcis}. 80\% of the dataset was randomly selected for model selection and building, and 20\% for model prediction on unseen sentences. To select
the optimal hyper parameter for each model, we performed stratified 10-fold cross-validation. Stratified sampling ensures that the proportion of each source would be kept. We ran our approach (e.g., KGConfig construction, security configuration document estimation, and security configuration concept classification) with different ML-models five times to calculate the time required to build KGSecConfig. Moreover, one-tailed non-parametric Mann-Whitney U-test \cite{manwhit} and Cohen's $d$ effect size \cite{cohend} tests were calculated to compare the statistical significance of the observed samples. These tests were selected since those were not affected by the data distribution and also used in prior studies to confirm their observation on the selection of ML-models \cite{puminer,puminer-15}.
\subsubsection{RQ3} We used a statistical sampling method similar to RQ1 to examine the randomly sampled instances of entities in KGSecConfig. Our statistical sampling ensures the evaluation is in 0.05 error margin at 95\% confidence level. We used accuracy as the evaluation metric for assessing the quality of KGSecConfig since  accuracy is the most used and well-known evaluation metrics for the quality assessment of KG as reported in the prior researches  \cite{332818, knowhow, bfskg, 332818_1}.  Two authors independently evaluated the accuracy and discussed to reach the consensus where independent assessment were different. We followed the prior research \cite{332818, knowhow, 332818_1} criterion, that is whether an extracted instance is correct and meaningful (e.g., complete). We computed Cohen's Kappa \cite{cohenkappa} to measure the inter rater agreement. Our SeConfigKG is built upon KGConfig (Section \ref{sec: CKG}) by adding our three secured configuration concepts (Section \ref{sec: SCCC}) as \textit{`statement'}, \textit{`goal'}, and \textit{`action'} entities using learning models. As we already evaluated the learning models in RQ2, we did not repeat the evaluation of these entities in RQ3. We focused our quality assessment on other configuration entities, e.g., argument, option, types, default values, and description.

\vspace{- 5 pt}
\subsection{Evaluation Results}
\subsubsection{RQ1} We achieved an accuracy of 0.98 for our keyword-based method. Besides, we obtained 1.00, 0.95, and 0.97 precision, recall, and f1-score, respectively. Our approach could not identify the configuration documents where syntax were not preserved in the document, e.g., \textit{`admissionControl'}, \textit{`--azure-container-registry'}, or \textit{`show.hidden.metrics'} violating camel case, hyphenation, or dotted syntax. However, relaxing the syntax preservation causes generation of huge false positives, e.g., providing irrelevant documents as configuration documents. For example, the sentences `The azure container registry is Microsoft's own hosting platform for Docker images' \cite{Allows-Attackers-Brick}, or `How to Best Secure Azure Container Registry' \cite{Secure-Your-Azure} resulted in a configuration sentence. However, the documents are not providing any configuration information. Thus, the relaxation of syntax preservation may generate many irrelevant documents to process and large search space. In this regard, our keyword-based method with syntax preservation can keep the search space of configuration knowledge localization minimized, yet representative and provide high precision results for retrieving secured configuration documents.    
\begin{figure}[]
  \centering
  \includegraphics[scale = 0.35]{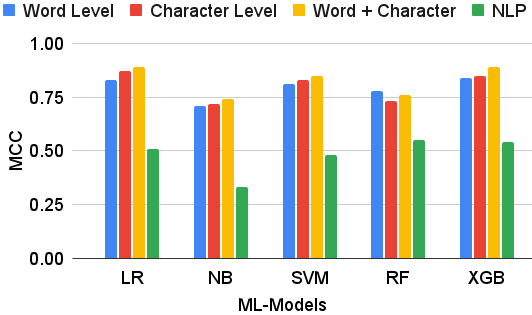}
  \caption{MCC for each of the ML-models and features}
  \label{fig: mcc}
\end{figure}

\begin{table}[]
\caption{Optimal hyper parameters for the studied ML-models}
\label{tab: hyper}
\scalebox{0.8}{
\begin{tabular}{|cccccc|}
\hline
Model & Hyper Parameter                                                                                                                                                                   & Model & Hyper Parameter                                                                 & Model & Hyper Parameter                                                                                                        \\ \hline
LR    & \begin{tabular}[c]{@{}c@{}}C: 2.60\\ fit\_intercept: True \\ max\_iter: 582 \\ solver: liblinear\\  tol: 9.95 e-05 \\ warm\_start: False \\ scale: 0 \\ normalize: 0\end{tabular} & SVM   & \begin{tabular}[c]{@{}c@{}}C: 8.68\\ gamma: 3.56 \\ kernel: linear\end{tabular} & RF    & \begin{tabular}[c]{@{}c@{}}criterion: gini \\ max\_depth: 34 \\ max\_features: auto \\ n\_estimators: 100\end{tabular} \\ \hline
NB    & \begin{tabular}[c]{@{}c@{}}alpha: 0.52 \\ fit\_prior: False\end{tabular}                                                                                                          & XGB   & \begin{tabular}[c]{@{}c@{}}max\_depth: 92\\ tree\_method: exact\end{tabular}    &       &     \\ \hline                                                                                                                  
\end{tabular}
}
\end{table}

\begin{table}[t]
\centering
\caption{Accuracy achieved by the studied ML-models}
\label{tab: accuracy}
\scalebox{0.8}{
\begin{tabular}{|c|c|c|c|c|c|}
\hline
Model & LR   & NB   & SVM  & RF   & XGB  \\ \hline
Accuracy & 0.94 & 0.82 & 0.88 & 0.76 & 0.93 \\ \hline
\end{tabular}
}
\end{table}

\begin{table*}[]
\caption{Precision, Recall, f1-score for each concept using the studied ML-models}
\label{tab: prec}
\scalebox{0.8}{
\begin{tabular}{c|ccc|ccc|ccc|ccc|ccc}
\multicolumn{1}{c}{} & \multicolumn{3}{c|}{LR}        & \multicolumn{3}{c|}{NB}        & \multicolumn{3}{c|}{SVM}       & \multicolumn{3}{c|}{RF}        & \multicolumn{3}{c}{XGB}       \\
\hline
concept              & Precision & Recall & f1-score & Precision & Recall & f1-score & Precision & Recall & f1-score & Precision & Recall & f1-score & Precision & Recall & f1-score \\
statement            & 0.98      & 0.93   & 0.96     & 0.94      & 0.77   & 0.85     & 0.93      & 0.93   & 0.93     & 0.97      & 0.82   & 0.89     & 0.97      & 0.9    & 0.93     \\
goal                 & 0.9       & 0.88   & 0.89     & 0.81      & 0.7    & 0.75     & 0.83      & 0.88   & 0.85     & 0.9       & 0.42   & 0.57     & 0.9       & 0.84   & 0.88     \\
action               & 0.86      & 0.93   & 0.9      & 0.7       & 0.91   & 0.79     & 0.88      & 0.83   & 0.85     & 0.58      & 0.96   & 0.72     & 0.86      & 0.91   & 0.89     \\
other                & 0.9       & 0.88   & 0.89     & 0.82      & 0.79   & 0.8      & 0.9       & 0.89   & 0.89     & 0.81      & 0.74   & 0.73     & 0.9       & 0.87   & 0.88  \\
\hline
\end{tabular}
}
\end{table*}
\subsubsection{RQ2} Fig. \ref{fig: mcc} shows a combination of word and character level feature performs better than other features in terms of MCC. For the character level feature, we chose $n$-gram in the range of 2 $\leq n$ $\leq$ 4 since vocabulary size does not increase after character size 4. A combination of word and character level features considers both word and character to learn the semantic representation of the sentence to give better results. NLP features (e.g., count of word, character, noun, verb, adjective, adverb, and pronoun) perform similarly for all single models except ensemble models. One of the reasons could be that ensemble models learn the overlapping of NLP features among classes due to their intrinsic characteristics of individual model aggregation. We verified NLP features perform statistically significant different than other features with  Mann-Whitney U-test ($z$-score is 2.506, $p$-value is .006, significant at $p$-value  $<$ .05) and large Cohen's \textit{d} effect size (4.324). Table \ref{tab: hyper} represents the optimal hyper-parameter of each model for its best feature, e.g., a combination of word and character level features. Besides, Table \ref{tab: accuracy} presents the accuracy achieved by different ML-model. It is observed that LR and XGB outperform other models. We verified our observation with  Mann-Whitney U-test ($z$-score is 3.741, $p$-value is 9 $\times$ 10$^-3$) and Cohen's \textit{d} effect size. Table \ref{tab: prec} shows the precision, recall, and f1-score obtained for each model. We found that LR and XGB also perform better than other models in terms of precision, recall, and f1-score. 

Table \ref{tab: time} shows the average time requirement to build KGSecConfig with different ML models. Our approach with the LR model takes the least amount of time (4.4 minutes on average) to build KGSecConfig. The difference among the time requirements is due to the different training times for ML models since the KGConfig construction, configuration documents estimation, and BFS algorithm execution required similar time to generate their respective output. XGB and RF take longer time to train than LR, NB, and SVM. We verified our observation with Mann-Whitney U-test, where we obtained statistical significant difference ($z$-score$<$3.185, $p$-value$<$9.1 $\times$10$^-3$, significant at $p$-value $<$ 0.05) and larger Cohen's $d$ effect size ($d>$174.1). XGB and RF are tree-based methods and traditionally reported to take longer time \cite{ganaie2021ensemble}. 

Calero and Pattini \cite{calero} mentioned current commercial designs are motivated by arguments based on sustainability (i.e., using fewer resources to achieve results). In particular, they asserted organizations used sustainability-based redesigns to motivate extensive cost-cutting opportunities. Therefore, we suggest LR should be preferred in building KGSecConfig.



\begin{table}[]
\centering
\caption{Time (s) required to build KGSecConfig by ML-models}
\label{tab: time}
\scalebox{0.8}{
\begin{tabular}{|c|c|c|c|c|c|}
\hline
Model & LR  & NB  & SVM & RF  & XGB \\ \hline
Time (in seconds) & 264 & 313 & 438 & 773 & 694 \\ \hline
\end{tabular}
}
\end{table}
\subsubsection{RQ3} Table \ref{tab: rq3} shows the results of our quality assessment of the extracted knowledge, where ACC-1, ACC-2, and ACC-F denotes the accuracy by two annotators independently, and final accuracy after resolving disagreement. The accuracies for all the configuration entities are above 90\% and Cohen's Kappa are above 0.6, indicating substantial to almost-perfect agreement between annotators.  

Our configuration entities argument and options achieved 100\% accuracy, which is not surprising since such entities are extracted from structured document contents (e.g., enclosed in html, xml tags) with careful implementation and verification. Null values, empty fields, and empty strings were the common problems for default values and types extraction. For example, \textit{`--allow-metric-labels'} argument in kube api server provided a square bracket notation to present empty fields in its default values, whereas \textit{`--cgroup-root'} provided double quotes, \textit{`--system-reserved'} provided backslash, and \textit{`--kube-reserved-cgroup'} kept the space blank after mentioning types (e.g., `--kube-reserved-cgroup', `Type:'   $<$space$>$) in Kubernetes, which  caused erroneous extraction. Besides, incomplete sentences due to erroneous splitting is the common problem for the description entity. For example,  \textit{`--container-runtime'} argument had the description `The container, e.g., docker, remote runtime to use'. Our sentence tokenizer based on NLTK provided two separate sentences as `The container, e.g.,', and `docker, remote runtime to use' causing incomplete sentences for the description entity. 

Our KGSecConfig contains highly accurate configuration knowledge, which can support practical downstream applications, such as automated misconfiguration detection, verification, and mitigation. The common problems with the quality of extracted knowledge  include text processing errors and meaningless description due to incomplete sentences. These problems can be minimized by developing more rules to enhance the text processing techniques, which can be further leveraged for  accurate configuration entity extraction.

\begin{table}[]
\centering
\caption{Accuracy of knowledge in KGSecConfig }
\label{tab: rq3}
\scalebox{0.8}{
\begin{tabular}
{|ccccc|} \hline 
Entities       & ACC-1 & ACC-2 & ACC-F & Agreement \\ \hline
argument       & 1.00  & 1.00  & 1.00  & 1.00      \\
options        & 1.00  & 1.00  & 1.00  & 1.00      \\
default values & 0.97  & 0.98  & 0.98  & 0.62      \\
type           & 0.96  & 0.94  & 0.94  & 0.77      \\
description    & 0.90  & 0.91  & 0.91  & 0.86    \\ \hline 
\end{tabular}}
\end{table}
\section{Implication and Discussion}
\vspace{- 5 pt}
\label{sec: discussion}

In this Section, we provide how KGSecConfig can be utilized for secured configuration of CO and some implications for researchers and practitioners.

\begin{figure}[]
  \centering
  \includegraphics[scale = 0.28]{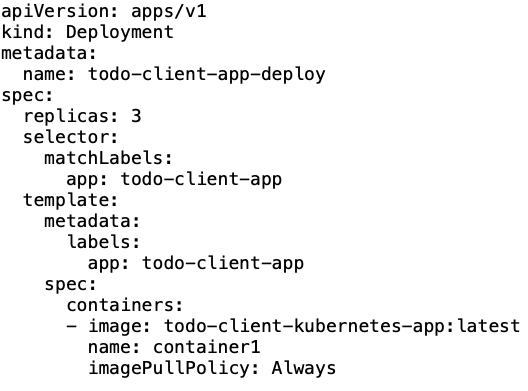}
  \caption{An example of Kubernetes manifest}
  \label{fig: misconfige}
\end{figure}

\begin{figure*}[]
  \centering
  \includegraphics[scale=0.28]{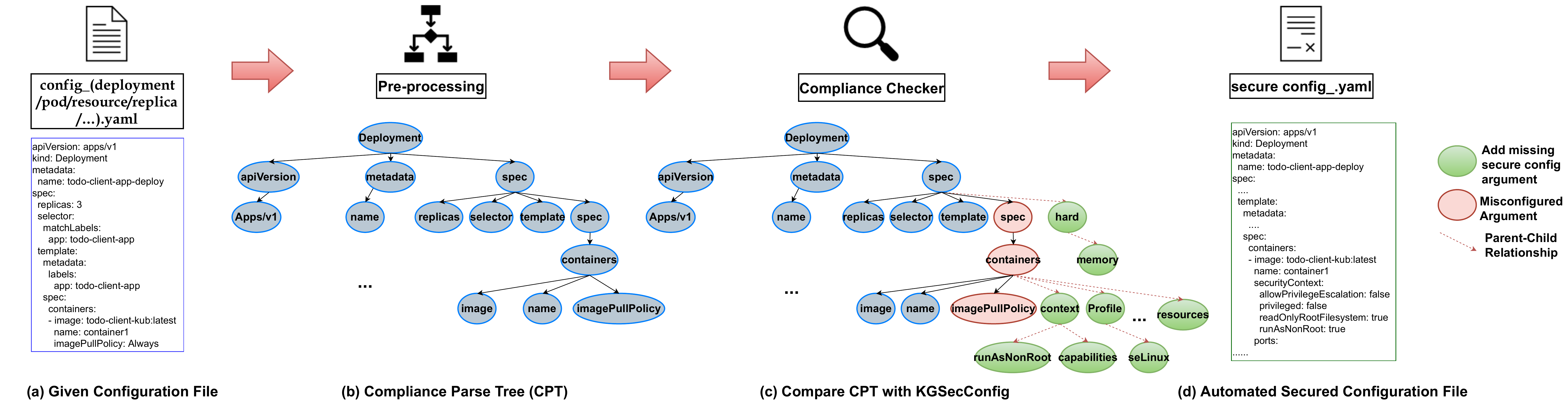}
  \caption{Automated misconfiguration mitigation using KGSecConfig}
  \label{fig: 3slides}
  
\end{figure*}
\vspace{- 2 pt}
\subsection{\textbf{Secured Configuration-as-a-Service}} 
\vspace{- 3 pt}
\subsubsection{Automated mitigation of misconfiguration} Our KGSecConfig can be used to mitigate the misconfiguration of CO clusters automatically. Fig. \ref{fig: misconfige} shows an example of a configuration file to deploy a containerized application \cite{github}. We built a Kubernetes cluster with two nodes \cite{nodes}, a control plane \cite{cplane}, and a worker node \cite{wnode} to explore the potential security threat actors with the configuration file as shown in Fig. \ref{fig: misconfige}. We leveraged our KGSecConfig to identify and mitigate the threat actors automatically, as shown in Fig. \ref{fig: 3slides}. A Compliance Parse Tree (CPT) as shown in Fig. \ref{fig: 3slides}(b) was built to tokenize configuration argument and its options, where the root of the tree is the configuration file type, e.g., \textit{`Deployment'}, non-leaf nodes are arguments, and leaf nodes are options. We developed a compliance checker using subgraph matching \cite{bfskg} whose aim is to compare CPT with our KGSecConfig for possible misconfiguration detection. 

Our KGSecConfig identified any user in the cluster could get access to the container as the security policy (e.g., \textit{`SecurityContext'}) argument is missing. A user can also access the container as a root user since \textit{`runAsNonRoot'} was not set. Moreover, a user can perform the privilege escalation as the \textit{`allowPrivilegeEscalation'} argument is missing. Besides, DDoS attack can be mounted due to configuring \textit{`imagePullPolicy'} as \textit{`Always'} and undefined host networking policy. All of the mentioned threat actors are automatically generated from our compliance checker, as shown in the red circle in Fig. \ref{fig: 3slides}(c). In particular, our \textit{`goal'} concept provided the knowledge of such threat actors where missing or undefined arguments were matched.  

Our KGSecConfig also has the capability to automatically mitigate the misconfiguration by replacing (e.g., replacing \textit{`Always'} option with \textit{`IfNotPresent'} in \textit{`imagePullPolicy'} argument) or adding (e.g.,  adding \textit{`allowPrivilegeEscalation'} argument with \textit{`false'} option) secured argument and corresponding options as a node or set of nodes as shown in green circles in Fig. \ref{fig: 3slides}(c) by using the knowledge in KGSecConfig. Finally, it can generate secured configuration files with YAML as shown in Fig. \ref{fig: 3slides}(d) or JSON format for automated execution due to the structured storage of relationships, e.g., parent-child relationship among arguments as shown in arrows in Fig. \ref{fig: 3slides}.

\subsubsection{Automated Verification of Extracted Configuration}
Our KGConfig (Section \ref{sec: CKG}) can serve as foundation support in terms of verification, e.g., whether the extracted configuration knowledge for security purposes either exists (e.g., if \textit{`--authorization-mode'} argument is consistent with the current documentation or \textit{`RBAC'} is a valid option for \textit{`--authorization-mode'}) according to official documentation. For instance, our KGSecConfig has detected some inconsistencies in the NSA published report \cite{CISA}. For instance, the NSA report stated Kubelet \cite{kubelet} and Kube-Scheduler \cite{kcs} run Transmission Control Protocol (TCP) on port number 10251, however, we identified it is actually 10250 for read/write port and 10255 read-only port. Similarly, the same inconsistencies were found in other components of Kubernetes, such as Kubelet \cite{kubelet} and control-manager \cite{kcm}. Our KGSecConfig can automatically detect inconsistencies without manual navigation of the large configuration space. In addition, we also detected some deprecated arguments (e.g., \textit{`--repair-malformed-updates'} in kube api-server) mentioned in some white papers (e.g., RedHat). Thus our KGSecConfig ensures real-time monitoring and updated configuration knowledge aligned with official documentation by automatic verification of extracted knowledge in terms of configuration space.     

\vspace{- 2 pt}
\subsection{\textbf{Automated Interpretation of Configuration Options}} 
\vspace{- 3 pt}
Our KGSecConfig can provide the reasoning why we need to set/enable or disable a particular set of arguments that can increase the understanding of the configuration knowledge. For example, \textit{`false'} option in \textit{`--profiling'} argument reduces the
potential attack surface since the default \textit{`true'} option generates a significant amount of program data that could potentially be exploited to uncover system and program details. Our KGSecConfig systematically organizes such essential knowledge by \textit{`hasGoal'} relationship with respective options which can help administrators' reasoning in configuring CO.  

Besides, our KGSecConfig minimizes the manual traversal of multiple data sources required for an investigation to change the configuration options due to the disclosure of new security issues, e.g., vulnerabilities or malware. For example, a denial-of-service vulnerability of kube api-server, CVE-2019-11254 is revealed in one security information sharing-source (i.e., in a vulnerability database NVD \cite{nvd}). This vulnerability can be fixed by restricting unauthorized access to kube api-server, which had been mentioned in another source, i.e., Security Announcement Kubernetes (SAK) \cite{kubernetes-security-announce}. However, neither NVD nor SAK described how to restrict unauthorized access to kube api server in terms of configuration options and arguments. Our KGSecConfig can correlate the encapsulated knowledge (e.g., \textit{`goal'} concepts associated with arguments of kube api server and the query `restrict unauthorized access to kube api-server') and return the mitigation approach by providing necessary arguments and options (e.g., \textit{`--anonymous-auth'} will be \textit{`false'} and \textit{`–authorization-mode'} will be \textit{`RBAC'}) for restricting the unauthorized access. Therefore, KGSecConfig can automatically provide the knowledge required for secured configuration to fix security issues. However, we acknowledge that the returned results will depend on the query formulation (e.g., query other than `restrict unauthorized access to kube api-server'). In future research, we will perform the embedding-based textual similarity \cite{sayagh} on the encapsulated knowledge and the given query to reduce the semantic and lexical gap.    
\vspace{- 1 pt}
\subsection{\textbf{Visualization of Configuration hot-spot}} 
\begin{figure}[]
  \centering
  \includegraphics[width = \linewidth]{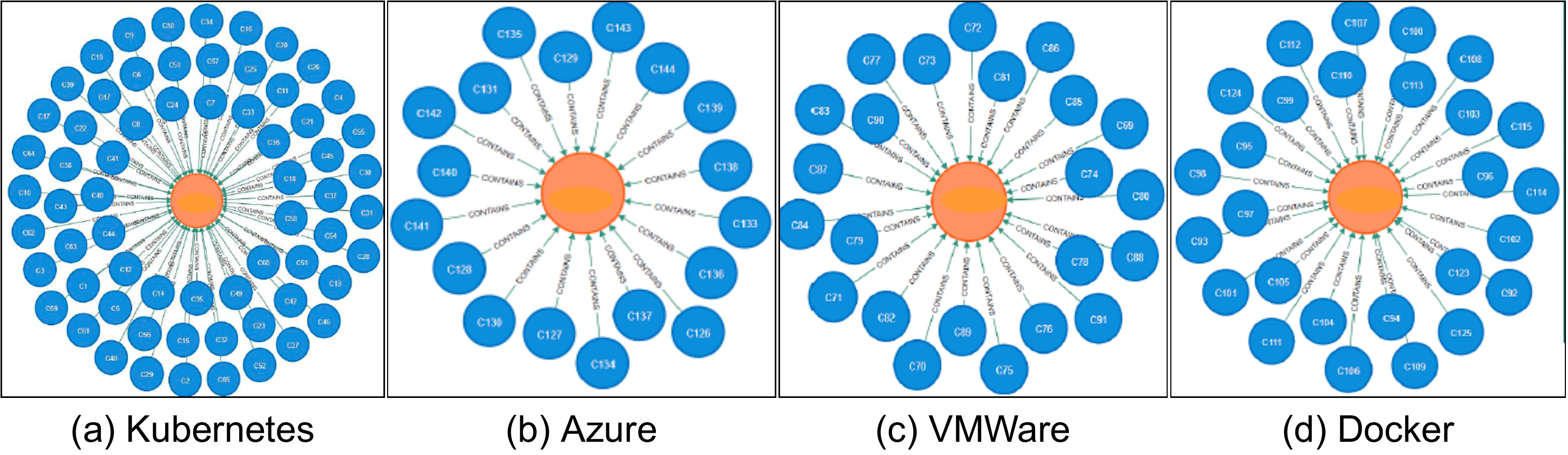}
  \caption{A subgraph visualizing the configuration hotspot of the studied CO software systems}
  \label{fig: vis}
\end{figure}

\vspace{- 3 pt} 
Our KGSecConfig can support to visualize the configuration hot-spot, i.e., the arguments which need to change their default options for security purposes. The default configurations in CO and its integrated software are not always security-focused and require changes to harden the CO. Multiple cyber-security attacks were launched due to default configuration in Kubernetes. For instance, Tesla went through a cyber breach because of misconfiguration in 2018 \cite{capitalonebreach}. Unauthorized users got access to secret resources exploiting an administrative CO which was being operated with default configuration in Tesla. Besides, the diversity of tools, platforms, and resource providers' configuration space makes it challenging to locate and then update the configuration manually. For example, the lead engineer of Target \cite{target}, a retail corporation, reviewed Kubernetes as ``..no document versioning. Stuff is all over. It is difficult to find the right stuff..'' \cite{trustradius}. In this regard, our KGSecConfig can be a potential approach to mitigate this challenge by visualizing a unified knowledge base. Fig. \ref{fig: vis} represents a sub-graph (built using Neo4j \cite{Neo4j}) providing the configuration hot-spot of the studied software systems.  Fig. \ref{fig: vis} can be a way to represent which software system needs to adjust default arguments more than the other software systems and can help the practitioner to prioritize their security configuration task. For instance, Fig. \ref{fig: vis}(a) has more more nodes compared to others (Fig. \ref{fig: vis}(b), \ref{fig: vis}(c), and \ref{fig: vis}(d)), representing it requires more adjustment of default arguments. Moreover, our KGSecConfig summarises relationships effectively (RQ1) and efficiently (RQ2 and RQ3), and it can scale to adopt new knowledge automatically using the learning-based model. 

\section{Threats to validity}
 \vspace{- 2 pt}
\label{sec: threats}
\subsubsection{Construct Validity} Labelled data are necessary for training a supervised learning model. Our manually labelled dataset may be biased and subjective. We followed the prior studies \cite{malkg,piplai,chadni} to mitigate the bias. Two authors independently annotated the dataset and reported the disagreement. Moreover, we only used the dataset that both authors agreed to train the learning model for minimizing the effect of bias. Our labelled dataset is also comparable with prior studies to build the cyber security knowledge graph. For example, Islam et al. \cite{chadni} labelled 1.7K, Rastogi et al. \cite{malkg} labelled 3K, Piplai et al. \cite{piplai} labelled 3.6K sentences for building security tools, predicting malware entities, and constructing malware KG, respectively.       
  
\subsubsection{Internal Validity}
Our collected configuration options from official documentation might be incomplete. However, prior studies on configuration mining from question/answering sites \cite{sayagh}, bug reports \cite{jin,wen} also leveraged official documentation in their studies. Besides, our security configuration arguments are derived from publicly available security information-sharing sources, which might miss some secured configuration arguments. However, we used security information sharing sources reported by prior researchers in the security of orchestrators \cite{shamim2020xi} or software systems. In future work, we will investigate other sources of security information. Another threat to the internal validity is the subjective judgment
in RQs, for example the evaluation of the quality of extracted configuration knowledge. To alleviate this threat, we have reported
the agreement for each subjective judgment.
    
\subsubsection{External Validity} Our evaluation results consider four case-study software systems and five ML models. We can not generalize our findings to other software systems or ML models. However, we provided an approach that can be applied to any other software system. Besides, our selected software system has a large configuration space and is popular in orchestration services \cite{kubeblog}. Moreover, our goal is not to compare different ML models, and any learning-based model can be plugged into our approach to build the KG. 

 
\vspace{- 6 pt}
\section{Conclusion and future work}
\vspace{- 5 pt}
\label{sec: conclussion}
85\% of global organizations are forecast to relocate their legacy application to container-based development and deployment \cite{stackrox}. Container orchestrators, e.g., Kubernetes, are playing a vital role in managing container clusters in both the development and production environment. Container orchestrators require automated configuration for secured operation of container clusters. The diversity of software systems used in orchestrators, their vast configuration space, and information overload cause barriers for automating the secured configuration. We proposed a novel knowledge graph-based approach, KGSecConfig to aggregate and organize disparate silos of security configuration knowledge.
We built a secured configuration knowledge graph with Kubernetes, Docker, Azure, and VMWare using KGSecConfig, which provide essential configuration knowledge, such as implementation details and reasoning. We demonstrated how KGSecConfig can be utilized for various downstream task, such as automated misconfiguration mitigation, inconsistency identification, and configuration hot-spot visualization. We plan to extend our KGSecConfig to explore the potential of secure configuration migration from one orchestrator to another orchestrator for the clusters. 

\section*{Acknowledgment}
The work has been supported by the Cyber Security Research Centre Limited whose activities are partially funded by the Australian Government’s Cooperative Research Centres Programme.
This work has also been supported with super-computing resources provided by the Phoenix HPC service at The University of Adelaide.    


\bibliographystyle{IEEEtran}
\bibliography{sample-base}

\end{document}